\documentclass[10pt,
aps,
 prl,
 twocolumn,
 superscriptaddress,
 amsmath,amssymb,
 altaffilletter,
 floatfix,
nofootinbib,
]{revtex4-2}

\usepackage{graphicx}\usepackage{dcolumn}\usepackage{bm}\usepackage{amsmath,amssymb}
\usepackage{xspace}
\usepackage{xcolor}
\usepackage{textcomp,gensymb}
\usepackage{multirow}
\usepackage{orcidlink}
\usepackage{array}
\newcolumntype{P}[1]{>{\centering\arraybackslash}p{#1}}

\usepackage[mathlines]{lineno}

\begin{document}
\setlength{\abovedisplayskip}{5pt}\setlength{\belowdisplayskip}{5pt}

\setlength{\abovecaptionskip}{1pt} \setlength{\belowcaptionskip}{0pt}

\newcommand{\gm}{\ensuremath{g\!-\!2}\xspace}

\newcommand{\runtwo}{Run-2\xspace}
\newcommand{\runthree}{Run-3\xspace}
\newcommand{\runthreea}{Run-3a\xspace}
\newcommand{\runthreeb}{Run-3b\xspace}
\newcommand{\runtwothree}{Run-2/3\xspace}
\newcommand{\RunFourFiveSix}{Run-4/5/6\xspace}

\newcommand{\omegaa}{$\omega_{a}$\xspace}
\newcommand{\omegaeta}{$\omega_{\eta}$\xspace}
\newcommand{\Aedm}{$A_{\mathrm{EDM}}$\xspace}
\newcommand{\Ag}{$C_{a}$\xspace}
\newcommand{\Sa}{$S_{a}$\xspace}
\newcommand{\Atilt}{$A_{\delta}$\xspace}

\newcommand{\Rfac}{$R$-factor\xspace}
\newcommand{\Rfacs}{$R$-factors\xspace}
\newcommand{\Rgam}{$R_{\gamma}$\xspace}
\newcommand{\Rpol}{$R_P$\xspace}
\newcommand{\Rposi}{$R_{e^+}(\lambda)$\xspace}
\newcommand{\Racc}{$R_{\mathrm{acc}}(\lambda)$\xspace}

\newcommand{\ang}{$\theta_y$\xspace}
\newcommand{\aveAng}{$\langle\theta_y\rangle$\xspace}
\newcommand{\aveAngTime}{$\langle\theta_y\rangle(t)$\xspace}

\newcommand{\note}[1]{\textcolor{red}{#1}\xspace}

\newcommand\Tstrut{\rule{0pt}{2.6ex}}         \newcommand\Bstrut{\rule[-0.9ex]{0pt}{0pt}}   %
 
\title{An improved direct limit on the muon electric dipole moment} 

\renewcommand{\thefootnote}{\fnsymbol{footnote}}
\affiliation{Argonne National Laboratory, Lemont, Illinois, USA}
\affiliation{Boston University, Boston, Massachusetts, USA}
\affiliation{Brookhaven National Laboratory, Upton, New York, USA}
\affiliation{Budker Institute of Nuclear Physics, Novosibirsk, Russia}
\affiliation{Center for Axion and Precision Physics (CAPP) / Institute for Basic Science (IBS)}
\affiliation{Cornell University, Ithaca, New York, USA}
\affiliation{Fermi National Accelerator Laboratory, Batavia, Illinois, USA}
\affiliation{INFN, Laboratori Nazionali di Frascati, Frascati, Italy}
\affiliation{INFN, Sezione di Napoli, Naples, Italy}
\affiliation{INFN, Sezione di Pisa, Pisa, Italy}
\affiliation{INFN, Sezione di Roma Tor Vergata, Rome, Italy}
\affiliation{INFN, Sezione di Trieste, Trieste, Italy}
\affiliation{Department of Physics and Astronomy, James Madison University, Harrisonburg, Virginia, USA}
\affiliation{Institute of Physics and Cluster of Excellence PRISMA++, Johannes Gutenberg University Mainz, Mainz, Germany}
\affiliation{Korea Advanced Institute of Science and Technology (KAIST)}
\affiliation{Michigan State University, East Lansing, Michigan, USA}
\affiliation{North Central College, Naperville, Illinois, USA}
\affiliation{Northern Illinois University, DeKalb, Illinois, USA}
\affiliation{Regis University, Denver, Colorado, USA}
\affiliation{School of Physics and Astronomy, Shanghai Jiao Tong University, Shanghai, China}
\affiliation{Tsung-Dao Lee Institute, Shanghai Jiao Tong University, Shanghai, China}
\affiliation{Department of Physics and Astronomy, Trinity University, San Antonio, Texas, USA}
\affiliation{Institut f\"ur Kern- und Teilchenphysik, Technische Universit\"at Dresden, Dresden, Germany}
\affiliation{Universit\`a del Molise, Campobasso, Italy}
\affiliation{Universit\`a di Udine, Udine, Italy}
\affiliation{Department of Physics and Astronomy, University College London, London, United Kingdom}
\affiliation{University of Illinois at Urbana-Champaign, Urbana, Illinois, USA}
\affiliation{University of Kentucky, Lexington, Kentucky, USA}
\affiliation{University of Liverpool, Liverpool, United Kingdom}
\affiliation{Department of Physics and Astronomy, University of Manchester, Manchester, United Kingdom}
\affiliation{Department of Physics, University of Massachusetts, Amherst, Massachusetts, USA}
\affiliation{University of Michigan, Ann Arbor, Michigan, USA}
\affiliation{University of Mississippi, University, Mississippi, USA}
\affiliation{University of Virginia, Charlottesville, Virginia, USA}
\affiliation{University of Washington, Seattle, Washington, USA}
\affiliation{City University of New York at York College, Jamaica, New York, USA}
\author{D.~P.~Aguillard\orcidlink{0000-0002-5556-1169}} \affiliation{University of Michigan, Ann Arbor, Michigan, USA}
\author{T.~Albahri\orcidlink{0000-0002-1191-5463}} \affiliation{University of Liverpool, Liverpool, United Kingdom}
\author{D.~Allspach\orcidlink{0009-0008-2417-5338}} \affiliation{Fermi National Accelerator Laboratory, Batavia, Illinois, USA}
\author{J.~Annala} \affiliation{Fermi National Accelerator Laboratory, Batavia, Illinois, USA}
\author{K.~Badgley\orcidlink{0000-0002-3942-5108}} \affiliation{Fermi National Accelerator Laboratory, Batavia, Illinois, USA}
\author{S.~Bae{\ss}ler\orcidlink{0000-0001-7732-9873}} \affiliation{University of Virginia, Charlottesville, Virginia, USA}
\author{L.~Bailey\orcidlink{0009-0004-7290-8963}} \affiliation{Department of Physics and Astronomy, University College London, London, United Kingdom}
\author{E.~Barlas-Yucel\orcidlink{0000-0001-9562-2125}} \altaffiliation[Now at ]{Fermi National Accelerator Laboratory, Batavia, Illinois, USA.} \affiliation{University of Illinois at Urbana-Champaign, Urbana, Illinois, USA}
\author{T.~Barrett\orcidlink{0000-0002-0807-5846}} \affiliation{Cornell University, Ithaca, New York, USA}
\author{E.~Barzi\orcidlink{0000-0001-5829-2147}} \affiliation{Fermi National Accelerator Laboratory, Batavia, Illinois, USA}
% ...
\renewcommand{\thefootnote}{\fnsymbol{footnote}}
\footnotetext[2]{Deceased.}
\renewcommand{\thefootnote}{\arabic{footnote}}
\collaboration{The Muon \gm Collaboration} \noaffiliation
\vskip 0.25cm

\date{\today}

\begin{abstract}
A limit on the permanent electric dipole moment (EDM) of the positive muon is presented based on data from the Fermilab Muon $g-2$ Experiment taken between 2019 and 2020.
The tracking detectors measure the average vertical decay angle of positrons from muon decays, enabling a search 
for an interaction between a possible muon EDM $d_\mu$ and the lab-frame magnetic field. The result, $d_\mu = (-0.35 \pm 0.19_{\mathrm{stat}} \pm 0.34_{\mathrm{sys}}) \times10^{-19}~e\cdot$cm, is consistent with zero and sets a new direct limit on the muon EDM of $|d_\mu|<1.10\times10^{-19}~e\cdot$cm at the 95\% confidence level.
\end{abstract} 
\maketitle

\textit{Introduction}---The known sources of charge-parity (CP) violation within the Standard Model (SM) are not sufficient to explain the observed matter-antimatter asymmetry of the universe \cite{PhysRevD.110.030001,Sakharov:1967dj}. Electric dipole moments (EDMs) violate parity and time-reversal (T) symmetries, so under the assumption of CPT conservation, EDMs provide an additional source of CP violation. 
So far, no EDM has been observed for a fundamental particle or for a bound state such as the neutron or various nuclei~\cite{electronEDM2023, PhysRevLett.124.081803,RevModPhys.91.015001}.

The EDM $\vec{d}$ and magnetic dipole moment $\vec{\mu}$ of a spin-1/2 particle can be written as
\begin{align}
\label{eq:eta}
  \vec{d}=\eta\frac{q}{2mc}\vec{s}, &&  \vec{\mu}=g\frac{q}{2m}\vec{s},
\end{align}
where $q$ is the charge, $m$ is the mass, $c$ is the speed of light, and $\vec{s}$ is the spin vector. The dimensionless constants $\eta$ and $g$ are analogous and quantify the coupling of the electric or magnetic field to the spin.
Muon-specific values of these constants are labeled in this paper as $d_{\mu}=\|\vec{d}_\mu\|$ and $\eta_{\mu}$.

The SM prediction for the muon EDM, 
generated by the hadron level long-distance effect,
is $\mathcal{O}(10^{-38})~e\cdot$cm  \cite{muonEDM_limits_small2020}.
An indirect limit can be obtained via a mass scaling of the electron EDM limit; using the 90\% CL in \cite{electronEDM2023} gives $|d_\mu|<8.5\times10^{-28}~e\cdot$cm. However, this assumes minimal flavor violation, which does not necessarily hold in beyond the SM (BSM) physics~\cite{Chivukula:1987fw, Hall:1990ac, Buras:2000dm, DAmbrosio:2002vsn}.
Another indirect limit has been calculated from muon-loop-induced interactions in nuclei giving $|d_\mu|<1.9\times10^{-20}~e\cdot$cm~\cite{Ema2022indirect}, although this exclusion is model dependent. 
The previous best direct limit from the Brookhaven National Laboratory $g-2$ experiment (BNL E821) of $|d_\mu|<1.8\times10^{-19}~e\cdot$cm at 95\% confidence level (CL) \cite{BNLEDM2009} is $\sim$19 orders of magnitude larger than the SM prediction. A measurement of a muon EDM inconsistent with zero near the current limits would disagree with the SM and inform BSM theories, e.g., \cite{DMedm_2023,Dekens_LQ_2019}.

There are multiple strategies to search for a muon EDM in a storage ring experiment where spin precession is described by Eqs.~\eqref{eq:full_omega_a} and \eqref{eq:omega_eta}.
One method measures the time variation in the average vertical angle of positrons from muon decays (the traceback method), while another looks at variations in the precession phase as a function of vertical position (the phase method).
The BNL E821 experiment used both methods, with the phase method being the most precise, though systematically limited \cite{BNLEDM2009}.
In contrast, at Fermilab, the traceback method has the highest sensitivity due to improved statistics, a new straw tracker system with significant improvements~\cite{FNAL_tracker_2022}, and refined analysis methods.
This analysis utilizes data from 2019--20, which divides into three datasets based on running conditions as described in \cite{FNAL_PRL_2023}. They are labeled \runtwo, \runthreea, and \runthreeb. 
\textit{Experiment}---A beam of highly polarized positive muons is injected into a 7.11~m radius, 1.45~T superferric storage ring magnet, with a design storage momentum of $3.094$~GeV/c. Magnetic kickers move the muon beam onto the central orbit just after injection and electrostatic quadrupoles (ESQs) provide weak vertical focusing. The 24 calorimeters and 2 straw tracker stations on the inside of the storage ring detect positrons from muon decays. Comprehensive details of the experiment have been set out previously in \cite{FNAL_TDR2018,FNAL_accel_2017,FNAL_calos2019,FNAL_PRL_2021,kickerpaper,FNAL_BeamDyn,FNAL_tracker_2022,FNAL_PRL_2023,FNAL_PRD_2024,FNAL_PRL_2025}. 

The EDM search uses data from the in-vacuum straw trackers.
There are two tracker stations around the storage ring: station 12 (S12) at $\sim180^\circ$ from the point of muon injection; and station 18 (S18) at $\sim270^\circ$. Positrons pass through the straws producing signals called hits that provide position measurements with an average resolution of $110~\micro$m. Track reconstruction proceeds by grouping hits using temporal and spatial information and then fitting. The track is then extrapolated back to estimate the muon decay position and time, and the initial positron 3-momentum. Full details of the trackers and track reconstruction can be found in \cite{FNAL_tracker_2022}.

\textit{Measurement principle}---Because $g_{\mu}>2$, the muon spin precesses ahead of the cyclotron frequency, with the anomalous spin precession frequency defined as the difference between these frequencies $\vec{\omega}_a = \vec{\omega}_s - \vec{\omega}_c$. In a storage ring with lab-frame electric field $\vec{E}$, magnetic field $\vec{B}$, and taking $d_\mu=0$, the anomalous precession frequency is
\thinmuskip=1.1mu \medmuskip=1.33mu \thickmuskip=1.7mu
\begin{equation}\label{eq:full_omega_a}
\begin{split}
    \vec{\omega}_a =& -a_\mu \frac{q}{m}\vec{B} \\
    &+ \frac{q}{m} \bigg[ \bigg(a_\mu-\frac{1}{\gamma^2-1}\bigg) \frac{\vec{\beta}\times \vec{E}}{c}
    - a_\mu \bigg(\frac{\gamma}{\gamma+1}\bigg)(\vec{\beta}\cdot\vec{B})\vec{\beta} \bigg],
\end{split}
\end{equation}
\thinmuskip=3mu \medmuskip=4mu \thickmuskip=5mu
where $a_\mu = (g_{\mu}-2)/2$ is the muon magnetic anomaly, $\vec{\beta}$ is the ratio of the velocity to the speed of light, and $\gamma$ is the Lorentz factor.
At the so-called `magic momentum' $p_0\approx3.094$~GeV/c, the term proportional to $\vec{E}$ vanishes. A cancellation of the third term occurs when there is no vertical betatron motion.

Due to the magnetic moment, a vertical magnetic field causes precession in the horizontal plane.
A non-zero EDM introduces a torque which causes the spin precession plane to be tilted. The extra contribution to the precession is 
\begin{equation}
\label{eq:omega_eta}
    \vec{\omega}_\eta = -\frac{q}{mc}\frac{\eta_\mu}{2} \left[ \vec{E} - \frac{\gamma}{1+\gamma}\left(\vec{\beta}\cdot\vec{E}\right)\vec{\beta} + c\vec{\beta}\times\vec{B} \right].
\end{equation}
With no transverse beam motion and neglecting the lab-frame electric field \footnote{When $\gamma=29.3$ and $|\vec{B}| = 1.45\mathrm{T}$, a lab-frame electric field of $|\vec{E}| \sim 435$~MV/m would provide the same contribution to $\omega_\eta$ as the $c\vec{\beta}\times\vec{B}$ term. The effect of the quadrupole electric field ($|\vec{E}| \sim 720$~kV/m at the plate boundaries) can therefore be neglected.}, this reduces to just the last term. 
The total precession frequency can then be written 
\begin{equation} \label{eq:omega_tot}
    \vec{\omega}_{\mathrm{TOT}}=\vec{\omega}_a + \vec{\omega}_\eta = - \frac{q}{m} \bigg[a_\mu\vec{B} + \frac{\eta_\mu}{2} \Big(\vec{\beta}\times\vec{B}\Big)\bigg].
\end{equation}
Assuming a vertical magnetic field and longitudinal momentum, the angle of the tilt in the spin precession plane is given by
\begin{equation}
    \delta=\tan^{-1}{\frac{\omega_\eta}{\omega_a}}\approx\frac{\omega_\eta}{\omega_a}=\frac{\beta\eta_\mu}{2a_\mu},
\end{equation}
where the small angle approximation is valid given $\omega_\eta\ll\omega_a$. 
Taking this with Eq.~\eqref{eq:eta} gives
\begin{equation} \label{eq:EDMtoTilt}
    d_\mu = \frac{q\hbar a_\mu}{2m_\mu c\beta}\delta,
\end{equation}
demonstrating that the EDM is proportional to the rest-frame tilt in the spin precession plane. 

Due to parity violation in the muon decay, the number of high energy decay positrons in the laboratory frame oscillates as the spin precesses. It is maximized when the spin and momentum vectors are aligned.
Any tilt in the precession plane will cause an oscillation in the average vertical decay angle $\langle\theta_y\rangle=\langle p_y/|p|\rangle$ which is maximized when the spin and momentum are orthogonal, contrary to the positron number oscillation. Therefore, this measurement looks for an oscillation in \aveAng that is $\pi/2$ out of phase with the anomalous spin precession, and at the same frequency.

\textit{Analysis method}---\ang is measured using momentum information from the straw trackers. The tracked positrons used in the analysis must meet a set of quality requirements to ensure accurate reconstruction, and must be detected in a time window of $(26.2\leq t\leq 602.4)$~\micro s after beam injection. 
The delayed start time avoids the high rates and unstable beam conditions immediately after injection, while maximizing statistics. Very few positrons are detected after the end time. The dataset used amounts to $2.34\times10^9$ positrons with 35\% from \runtwo, 47\% from \runthreea, and 18\% from \runthreeb.

As measured in the laboratory frame, the amplitude \Atilt of the oscillation in \aveAng from a tilted spin precession plane is reduced from the rest-frame tilt by several factors. These four reductions, or \Rfacs, are written as 
\begin{equation} \label{eq:delta_m_to_delta_MRF}
    A_{\delta} = R_\gamma R_P R_{e^+}(\lambda)R_{\mathrm{acc}}(\lambda)\delta,
\end{equation}
where $\lambda=E/E_{\mathrm{max}}\approx p_{e^+}/p_{\mathrm{max}}$, $R_\gamma=1/\gamma \approx 0.03$ is from the Lorentz boost, and \Rpol is the average beam polarization of 94\%.
The momentum dependent spread of decay angles relative to the muon spin vector reduces the measured average angle by the factor \Rposi, which depends on the positron momentum.
Simulation is used to determine this value~\cite{Geant42016}, which includes the effect of radiative corrections~\footnote{\texttt{GEANT4} includes $\mu^+\rightarrow e^+\nu_e\bar{\nu}_\mu\gamma$ for both initial and final state radiation.}.
The final reduction factor \Racc depends on momentum and encompasses the impact of detector acceptance. Due to the location and extent of the trackers, smaller angles are reconstructed with higher efficiency, which decreases the measured amplitude \Atilt. 
The calculation of \Racc takes the ratio of the measured amplitudes \Atilt in two simulated data samples generated with a nonzero EDM, one with realistic acceptance and the other with $100\%$ acceptance.
The simulation is tuned to the data to match the measured vertical beam position $y$ and $\theta_y$ averaged over a dataset.
Because the EDM sensitivity depends on momentum, the analysis is carried out in eight non-overlapping 250~MeV/c wide bins of positron momentum ($p_{e^+}$) in the range 750--2750~MeV/c. 
The values for the momentum-dependent reduction factors in these eight bins range from $0.05 <$ \Rposi $< 0.25$ and $0.3 <$  \Racc $< 0.8$.

$\theta_y$ is binned according to decay time modulo the precession period. Beyond the vertical angle oscillation due to an EDM, there are beam motions that, if not accounted for, may impact the sensitivity to an EDM.
To minimize the impact of oscillations at frequencies other than the signal frequency \omegaa, \aveAngTime is measured modulo the precession period $T_a=4.37$~\micro s. 
The number of bins was chosen to give a bin width ($150.5$~ns) as close as possible to the cyclotron period $T_c = 149.2$~ns, which reduces the effects of cyclotron period modulation \cite{FNAL_PRD_2024}.

A uniform randomization with temporal width equal to the vertical betatron (VB) period $T_y=0.45$~\micro s removes the oscillation at this frequency from the data. A correction is applied to account for an exponential decrease in \aveAng at early fit times when the beam position was stabilizing.

\textit{Fitting}---\aveAngTime is fitted with the following function,
\thinmuskip=2.7mu \medmuskip=3.6mu \thickmuskip=4.5mu
\begin{equation} \label{eq:fullEDMfit}
    \langle\theta_y\rangle(t) = \frac{C_{a}\cos(\omega_at+\phi_a)+S_{a}\sin(\omega_at+\phi_a)}{f(t)} +C.
\end{equation}
\thinmuskip=3mu \medmuskip=4mu \thickmuskip=5mu
\Sa accounts for any oscillation in \aveAng that is $\pi/2$ out of phase with the anomalous precession (from an EDM or EDM-like signal) and \Ag allows for any in-phase oscillations. The constant term $C$ allows for a non-zero average angle due to acceptance or beam position. The phase of spin precession relative to the muon momentum $\phi_a$ is fixed to the anomalous oscillation phase as fitted using data from the trackers with $p_{e^+}>1.7$~GeV/c. The frequency \omegaa is fixed to the value measured using the calorimeters in \cite{FNAL_PRL_2023}.
The denominator $f(t)$ is necessary to normalize to the number oscillation from the anomalous precession. It is given by
\begin{equation} \label{eq:denom}
\begin{split}
    f (t) = & \left(1+A\cos[\omega_a t+\phi^p_a]\right) \\
    & \times \left(1+A_{\mathrm{CBO}}\cos[\omega_{\mathrm{CBO}} t+\phi_{\mathrm{CBO}}]\right),
\end{split}
\end{equation}
where CBO refers to the coherent betatron motion of the beam, and the asymmetry $A$, momentum-dependent phase $\phi^p_a$, CBO amplitude $A_{\mathrm{CBO}}$, frequency $\omega_{\mathrm{CBO}}$, and phase $\phi_{\mathrm{CBO}}$ are obtained from fits to the number oscillation in each momentum bin. These fits use Eq.~\eqref{eq:denom} with the addition of two extra terms to account for the exponential decay at the boosted muon lifetime
~\footnote{$A_{\mathrm{CBO}}$ is also exponentially decreasing as given in \cite{FNAL_PRD_2024}, but given the long lifetime $\sim250$~\micro s it is excluded in the modulated fits.}.
The phase $\phi^p_a$ is momentum-dependent and distinct from the reference phase such that the normalization is correct in the given $p$ bin.
The $\theta_y$ fit fixes the $f(t)$ parameters.

\textit{Blinding}---The analysis was initially blinded by injecting an unknown oscillation at the EDM phase on top of the data before fitting.
The size of the blinding was determined by an unknown random number between 2 and 8 that was used as a multiplier on the BNL muon EDM limit.
This EDM is converted to a tilt via Eq.~\eqref{eq:EDMtoTilt}, then translated to \Sa using Eq.~\eqref{eq:delta_m_to_delta_MRF} and the relevant \Rfacs for the dataset.
The blinding is added to each data point using the function in Eq.~\eqref{eq:fullEDMfit} with \Ag$=0$, and $f(t)$ set from the anomalous precession fit. \runtwo and \runthree (the combination of \runthreea and \runthreeb) had separate blinding values until the necessary checks were complete whereupon a common blinding was used.

The unblinding proceeds in two phases. An initial \textit{soft} unblinding checked whether the measured muon EDM was discrepant from zero by more than three standard deviations, which would strongly contradict both SM predictions and indirect limits. The absolute size and sign of the measured signal are not revealed during this procedure. The analysis failed this initial \textit{soft} unblinding check. Following the predefined collaboration procedure, this prompted an internal review, leading to the discovery and characterization of the vertical misalignment and longitudinal magnetic field uncertainties described below, while the fitted data remained blinded. The former is a source of a large EDM-like signal, and a data-driven correction for this false signal was developed and applied.  
After the additional scrutiny and both the sign and size of the new correction were finalized, a full unblinding was performed, where the fits are repeated on data without an injected signal.

\begin{figure} [htp]
    \centering
        \centering
        \includegraphics[width=\columnwidth]{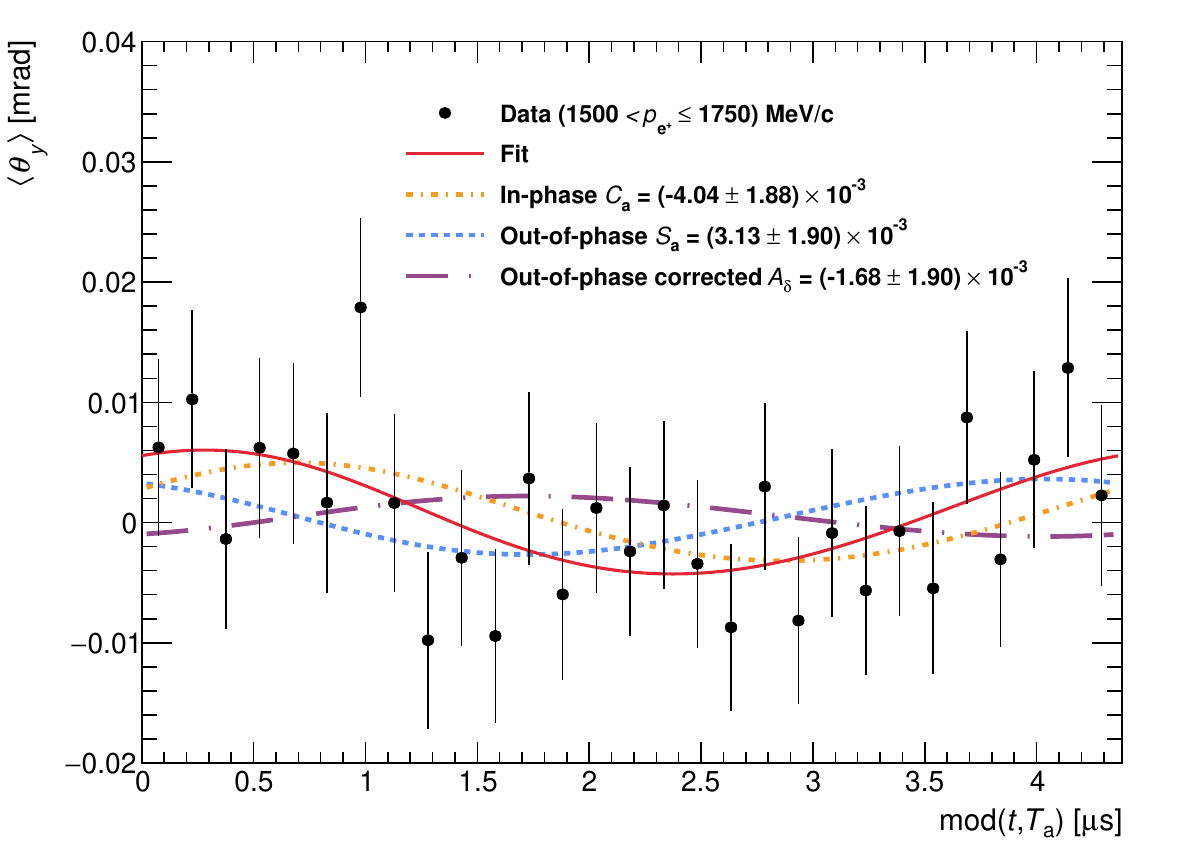}
        \caption{Fit of Eq.~\eqref{eq:fullEDMfit} to \aveAngTime for Run-3a S18 with $(1500<p_{e^+}\leq1750)$~MeV/c. The components of the fit in and out of phase with the anomalous precession (\Ag and \Sa respectively) are shown in orange and blue. After the alignment correction is applied, the purple curve shows the remaining out-of-phase component with amplitude \Atilt.}
        \label{fig:vangle_r3a_S18_1500_1750}
\end{figure}

\textit{EDM-like signals}---The EDM signature is a difference in the radially inward and outward decays, so other quantities that vary with decay angle are potential sources of fake signals. One such quantity is the path length from decay to detector, which is longer for outward decays, causing a widening of the vertical distribution. The tracker acceptance truncates the vertical position distribution and, when combined with a vertical misalignment of the trackers, changes the mean vertical position. Hence, the varying width causes the mean vertical position to vary out of phase with the anomalous spin precession. 
The large correlation between vertical position and vertical angle results in the measurement of non-zero \Ag and \Sa.
\Ag from this effect has a distinct momentum-dependent shape. 
Matching this shape in simulation to the data by offsetting the tracks determined the vertical misalignment. These offsets translated into a correction to remove the false EDM-like amplitude.
Fig.~\ref{fig:vangle_r3a_S18_1500_1750} shows the fitted \Ag, the measured \Sa and the out-of-phase amplitude after correction.

In addition, Eq.~\eqref{eq:omega_tot} shows that a radial component of the magnetic field ($B_r$) results in \omegaa having a component parallel to \omegaeta. This would tilt the spin precession plane out of the horizontal, providing a mechanism for an EDM-like signal. The rest-frame tilt due to a radial field $\delta_{B_r}$ is given by 
\begin{equation}
    \delta_{B_r}\ (\mathrm{\micro rad}) = \frac{\langle B_r\rangle}{B_y}\ (\mathrm{ppm}),
\end{equation}
where $\langle B_r\rangle$ is the average radial field around the ring and $B_y=1.45$~T is the vertical field component.
To determine this tilt, the radial field is measured by exploiting the relationship between $\langle B_r\rangle$, the vertical position of the beam, and the voltage of the ESQs. The induced rest-frame tilt is calculated for \runtwo, \runthreea, and \runthreeb separately and ranges from 11--15~\micro rad. These tilts are subtracted from the measured rest-frame tilt in each dataset before combining results.

Lastly, a longitudinal magnetic field with a first-order azimuthal harmonic will also induce an EDM-like signal that depends on the azimuthal position of the detector. One measurement of the longitudinal field around the ring was performed before the start of Run-1. These data are fitted to determine the $n=1$ component, and the maximum equivalent EDM for any detector position is taken as an uncertainty on $d_\mu$. 

\textit{Fit results}---
The data are fitted separately for each run, station, and momentum bin.
Fig.~\ref{fig:vangle_r3a_S18_1500_1750} shows the fit to \aveAng for the most sensitive momentum bin, with the out-of-phase and in-phase components of Eq.~\eqref{eq:fullEDMfit} plotted separately.
The reduced $\chi^2$ for each of the 48 fits vary between 0.6 -- 1.8, with an average value of 1.08.

\textit{Systematic uncertainties}---This section outlines how each systematic uncertainty is obtained. The uncertainties are summarized in Tab.~\ref{tab:sysSumm}.

Using different seeds for the vertical betatron randomization gives a spread in \Sa that is taken as a systematic uncertainty. After varying the early-time correction function within the parameters' uncertainties, a small uncertainty is determined from the subsequent shift in \Sa. The reference $g-2$ phase is measured with momentum selection varied between $(1700\pm100)$~MeV/c and the resulting shifts in phase used to assess the difference in \Sa. The uncertainty on each parameter in the $f(t)$ fit is propagated through to the fit to \aveAng, giving a subdominant systematic contribution. There is a spread in \Sa due to track reconstruction that is assessed with simulation. The alignment of the trackers in the reconstruction is varied within its uncertainty to assess the effect on \Sa. A small uncertainty is assigned to the potential for a false signal from beam-dynamics effects coupling to acceptance. These are all uncertainties on the measured \Sa and are converted to uncertainties on $d_\mu$ via Eqs.~\eqref{eq:EDMtoTilt} and \eqref{eq:delta_m_to_delta_MRF}. 

Each \Rfac has an associated relative uncertainty: \Rgam from the spread in muon momentum of 0.15\% RMS, \Rpol from a comparison with data and simulation of $2\%$,
and \Rposi from the statistical uncertainty on the simulation in each momentum bin $\mathcal{O}(0.5\%)$. The uncertainty on \Racc has contributions from simulation statistics and data and simulation matching giving $\mathcal{O}(5\%)$. The resulting uncertainties on $d_\mu$ are given via standard uncertainty propagation.

The dominant uncertainty arises from the correction for a false signal arising from vertical misalignment. The statistics of the data used for template matching drives the size of the uncertainty. Mismodeling in the simulation sample and its statistical uncertainty also contribute.
The radial magnetic field introduces a subdominant uncertainty on the rest-frame tilt, which is converted to an uncertainty on the EDM via Eq.~\eqref{eq:EDMtoTilt}. Similarly, the longitudinal field measurement adds an uncertainty on the measured EDM.
Any uncertainties on the fundamental constants in Eq.~\eqref{eq:EDMtoTilt} are negligible.

\begin{table}[htb]
    \caption{Contributions to the uncertainty on $d_\mu$ from statistical and systematic sources.}
    \centering
    \begin{ruledtabular}
    \begin{tabular}{lc}
    Uncertainty & \shortstack{$\sigma_{d_\mu}$ ($\times 10^{-20}~e\cdot$cm)}\Bstrut\\
    \hline
    Statistical & 1.87\Tstrut\\
    Total systematic & 3.43\Bstrut\\
    VB randomization & 0.39\Tstrut\\
    Early-time correction & 0.06\\
    $g-2$ phase & 0.14\\
    $f(t)$ fit & $<0.01$\\
    Track reconstruction & 0.59\\
    Reconstruction alignment & 0.86\\
    Beam dynamics & 0.89\\
    \Racc & 0.03\\
    \Rposi & 0.01\\
    \Rpol & 0.07\\
    \Rgam & 0.01\\
    Vertical alignment correction & 3.12\\
    Radial magnetic field & 0.02\\
    Longitudinal magnetic field & 0.33\Bstrut\\
    \end{tabular}
    \end{ruledtabular}
    \label{tab:sysSumm}
\end{table}

\textit{Combination}---Each measurement of \Sa is corrected to account for the alignment effect, then converted to a rest-frame tilt via Eq.~\eqref{eq:delta_m_to_delta_MRF} using the corresponding \Rfacs for the run, station, and momentum bin. The tilt due to the radial field is then subtracted. Finally, the tilt is converted to an EDM via Eq.~\eqref{eq:EDMtoTilt}. The EDM measurements are combined using a best linear unbiased estimator (BLUE) \cite{Lyons:1988rp}, to account for correlations between systematic uncertainties. 

\begin{figure} [htp]
    \centering
    \includegraphics[width=\columnwidth]{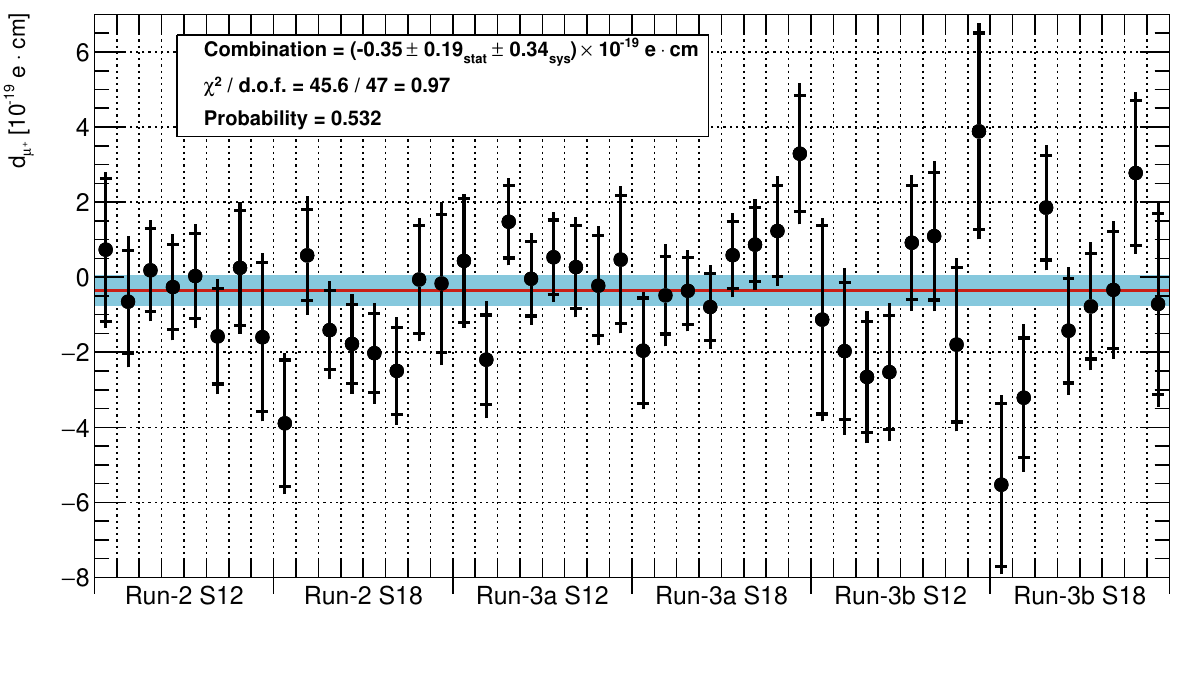}
    \caption{Summary of all 48 measurements. Each run and station has results from eight momentum bins shown in increasing $p_{e^+}$. The inner error bar is statistical and the outer includes systematic uncertainties. The result of the combination is shown in red, with its uncertainty in blue.}
    \label{fig:combo}
\end{figure}

Fig.~\ref{fig:combo} show the result of the combination, where the statistical uncertainty is from the statistical only combination. The systematic uncertainty values in Tab.~\ref{tab:sysSumm} are determined by repeating the combination without that systematic source and taking the quadrature difference in the uncertainty from the complete combination.

\textit{Cross-checks}---The combination is calculated in each station and run separately and no significant deviation across these parameters is found. The pull distribution from all datasets is consistent with a mean of 0 and standard deviation of 1.
Scans over fit start-time are performed for all datasets with the differences falling within the expected variation, which confirms that the chosen time does not introduce a bias. The analysis is also repeated using data from the total momentum range for each station and run; the results agree with the momentum-binned combination.

\textit{Measurement and limits}---
The final combination of all 48 measurements is shown in Fig.~\ref{fig:combo}, giving a muon EDM of
\begin{equation*}
    d_{\mu^+} = (-0.35 \pm 0.19_\text{stat} \pm 0.34_\text{sys}) \times 10^{-19}~e\cdot\text{cm}.
\end{equation*}
Given that this measurement is consistent with zero, it can be used to set an upper limit on the muon EDM.
Using the Feldman-Cousins limit setting procedure \cite{FeldmanCousins}, our result gives one-sided limits of
\begin{equation*}
    |d_{\mu}|<1.10\times10^{-19}~e\cdot\text{cm at 95\% CL}
\end{equation*}
and
\begin{equation*}
    |d_\mu|<0.98\times10^{-19}~e\cdot\text{cm at 90\% CL.}
\end{equation*}
Combining with the result from BNL of $(0.0\pm0.9)\times10^{-19}~e\cdot\text{cm}$, this becomes $|d_\mu|<0.98(0.86)\times10^{-19}~e\cdot$cm at the 95(90)\% CL.

\begin{figure} [htp]
    \centering
    \includegraphics[width=\columnwidth]{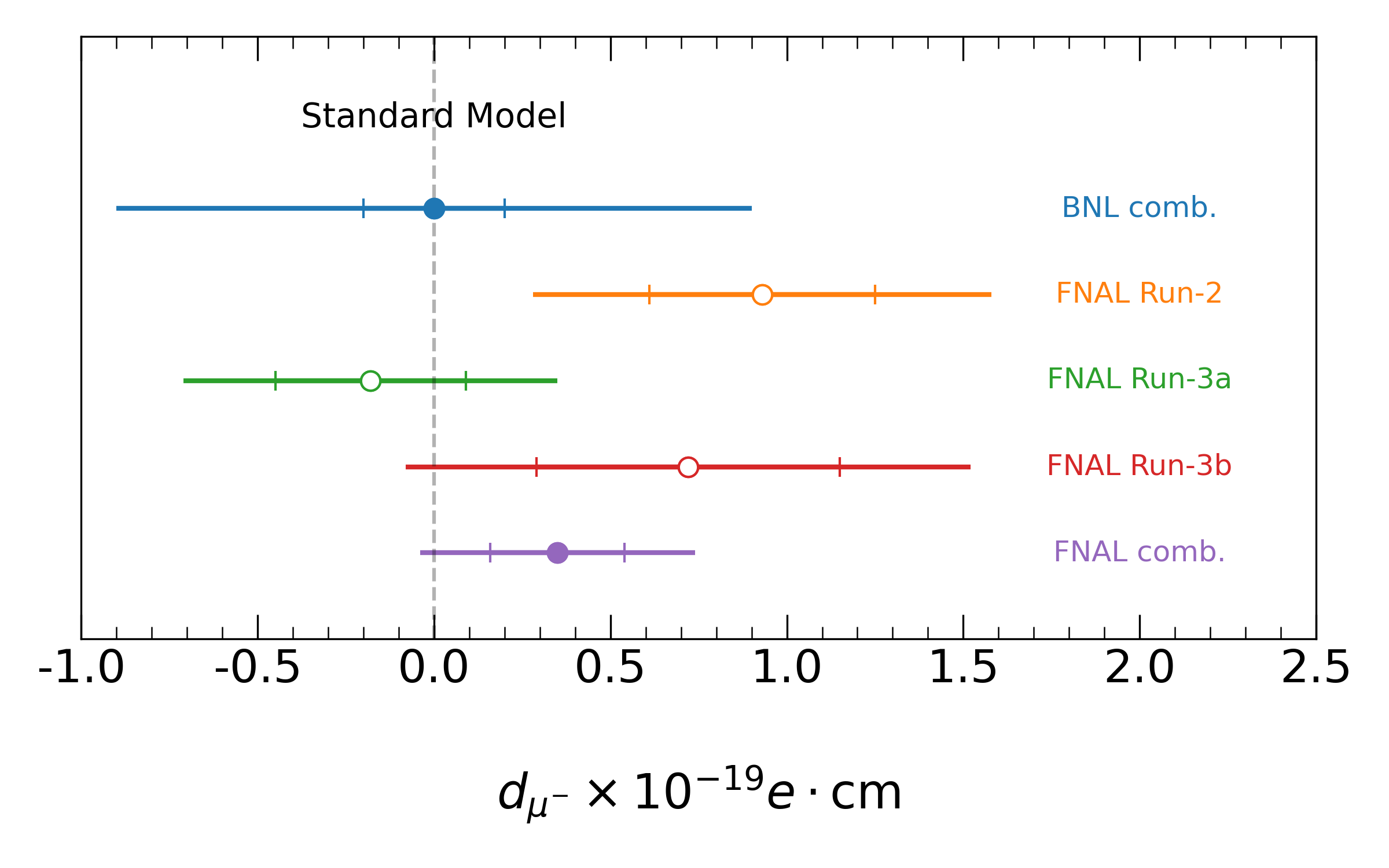}  
    \caption{The Fermilab muon EDM measurements for each dataset, and their combination, in comparison with the BNL measurement. The inner error bar for each data point is statistical and the outer includes systematic uncertainties. Each measurement has been recast in terms of the negative muon assuming CPT invariance. The SM expectation is also shown.}
    \label{fig:compare}
\end{figure}

A summary of the corresponding measurements for the negative muon, assuming CPT invariance, is shown in Fig.~\ref{fig:compare}. The ratio of the total uncertainties of the BNL and FNAL results, which gives an estimate of the relative sensitivity, is $2.3$. 

\textit{Discussion}---The result presented here sets a new direct limit on the muon EDM, improving on the previous 95\% C.L limit by a factor of 1.5. This is a systematically-limited result, with the largest contribution to the uncertainty coming from vertical alignment determination. We note this uncertainty itself is statistically limited and will improve when including the full dataset from Runs 4, 5, and 6, which together constitute 2.5 times the statistics presented here.
 
\textit{Acknowledgements}---We thank the Fermilab management and staff for their strong support of this experiment,
as well as our university and national laboratory engineers, technicians, and workshops
for their tremendous support.

The Muon \gm\ Experiment was performed at the Fermi National Accelerator Laboratory,
a U.S. Department of Energy, Office of Science, HEP User Facility. Fermilab is managed
by Fermi Forward Discovery Group, LLC, acting under Contract No. 89243023CSC000002.
Additional support for the experiment was provided by the U.S. DOE Office of Science
through the offices of HEP, NP, ASCR, by the U.S.-Japan Science and Technology
Cooperation Program in HEP, by the National Science Foundation (USA), by the Istituto
Nazionale di Fisica Nucleare (Italy), by the Science and Technology Facilities Council
(UK) and the Royal Society (UK) under Grant No.~URF$\backslash$R1$\backslash$231503, 
by the National Natural Science Foundation of China (Grant No. 12475108, 12305217, 12075151), 
by the Ministry of Science, ICT and Future Planning (MSIP), the National Research 
Foundation of Korea (NRF), and the Institute for Basic Science under Project Code 
No.~IBS-R017-D1 (Republic of Korea), by the German Research Foundation (DFG) 
through the Cluster of Excellence PRISMA+ (EXC 2118/1, Project ID 39083149) and the 
Cluster of Excellence PRISMA++ (EXC 2118/2, Project ID 39083149), by the European Union 
Horizon 2020 research and innovation programme under the Marie Sk\l{}odowska-Curie 
Grant Agreements No. 101006726 and No. 734303, by the European Union STRONG 2020 project 
under Grant Agreement No. 824093, and by the Leverhulme Trust, LIP-2021-014.


\begin{thebibliography}{27}\makeatletter
\providecommand \@ifxundefined [1]{\@ifx{#1\undefined}
}\providecommand \@ifnum [1]{\ifnum #1\expandafter \@firstoftwo
 \else \expandafter \@secondoftwo
 \fi
}\providecommand \@ifx [1]{\ifx #1\expandafter \@firstoftwo
 \else \expandafter \@secondoftwo
 \fi
}\providecommand \natexlab [1]{#1}\providecommand \enquote  [1]{``#1''}\providecommand \bibnamefont  [1]{#1}\providecommand \bibfnamefont [1]{#1}\providecommand \citenamefont [1]{#1}\providecommand \href@noop [0]{\@secondoftwo}\providecommand \href [0]{\begingroup \@sanitize@url \@href}\providecommand \@href[1]{\@@startlink{#1}\@@href}\providecommand \@@href[1]{\endgroup#1\@@endlink}\providecommand \@sanitize@url [0]{\catcode `\\12\catcode `\$12\catcode
  `\&12\catcode `\#12\catcode `\^12\catcode `\_12\catcode `\%12\relax}\providecommand \@@startlink[1]{}\providecommand \@@endlink[0]{}\providecommand \url  [0]{\begingroup\@sanitize@url \@url }\providecommand \@url [1]{\endgroup\@href {#1}{\urlprefix }}\providecommand \urlprefix  [0]{URL }\providecommand \Eprint [0]{\href }\providecommand \doibase [0]{https://doi.org/}\providecommand \selectlanguage [0]{\@gobble}\providecommand \bibinfo  [0]{\@secondoftwo}\providecommand \bibfield  [0]{\@secondoftwo}\providecommand \translation [1]{[#1]}\providecommand \BibitemOpen [0]{}\providecommand \bibitemStop [0]{}\providecommand \bibitemNoStop [0]{.\EOS\space}\providecommand \EOS [0]{\spacefactor3000\relax}\providecommand \BibitemShut  [1]{\csname bibitem#1\endcsname}\let\auto@bib@innerbib\@empty
\bibitem [{\citenamefont {Navas}\ \emph {et~al.}(2024)\citenamefont {Navas}
  \emph {et~al.}}]{PhysRevD.110.030001}\BibitemOpen
  \bibfield  {author} {\bibinfo {author} {\bibfnamefont {S.}~\bibnamefont
  {Navas}} \emph {et~al.} (\bibinfo {collaboration} {Particle Data Group
  Collaboration}),\ }\bibfield  {title} {\bibinfo {title} {Review of particle
  physics},\ }\href {https://doi.org/10.1103/PhysRevD.110.030001} {\bibfield
  {journal} {\bibinfo  {journal} {Phys. Rev. D}\ }\textbf {\bibinfo {volume}
  {110}},\ \bibinfo {pages} {030001} (\bibinfo {year} {2024})}\BibitemShut
  {NoStop}\bibitem [{\citenamefont {Sakharov}(1967)}]{Sakharov:1967dj}\BibitemOpen
  \bibfield  {author} {\bibinfo {author} {\bibfnamefont {A.~D.}\ \bibnamefont
  {Sakharov}},\ }\bibfield  {title} {\bibinfo {title} {{Violation of CP
  Invariance, C asymmetry, and baryon asymmetry of the universe}},\ }\href
  {https://doi.org/10.1070/PU1991v034n05ABEH002497} {\bibfield  {journal}
  {\bibinfo  {journal} {Pisma Zh. Eksp. Teor. Fiz.}\ }\textbf {\bibinfo
  {volume} {5}},\ \bibinfo {pages} {32} (\bibinfo {year} {1967})}\BibitemShut
  {NoStop}\bibitem [{\citenamefont {Roussy}\ \emph {et~al.}(2023)\citenamefont {Roussy}
  \emph {et~al.}}]{electronEDM2023}\BibitemOpen
  \bibfield  {author} {\bibinfo {author} {\bibfnamefont {T.~S.}\ \bibnamefont
  {Roussy}} \emph {et~al.},\ }\bibfield  {title} {\bibinfo {title} {An improved
  bound on the electron’s electric dipole moment},\ }\href
  {https://doi.org/10.1126/science.adg4084} {\bibfield  {journal} {\bibinfo
  {journal} {Science}\ }\textbf {\bibinfo {volume} {381}},\ \bibinfo {pages}
  {46–50} (\bibinfo {year} {2023})}\BibitemShut {NoStop}\bibitem [{\citenamefont {Abel}\ \emph {et~al.}(2020)\citenamefont {Abel} \emph
  {et~al.}}]{PhysRevLett.124.081803}\BibitemOpen
  \bibfield  {author} {\bibinfo {author} {\bibfnamefont {C.}~\bibnamefont
  {Abel}} \emph {et~al.},\ }\bibfield  {title} {\bibinfo {title} {Measurement
  of the permanent electric dipole moment of the neutron},\ }\href
  {https://doi.org/10.1103/PhysRevLett.124.081803} {\bibfield  {journal}
  {\bibinfo  {journal} {Phys. Rev. Lett.}\ }\textbf {\bibinfo {volume} {124}},\
  \bibinfo {pages} {081803} (\bibinfo {year} {2020})}\BibitemShut {NoStop}\bibitem [{\citenamefont {Chupp}\ \emph {et~al.}(2019)\citenamefont {Chupp},
  \citenamefont {Fierlinger}, \citenamefont {Ramsey-Musolf},\ and\
  \citenamefont {Singh}}]{RevModPhys.91.015001}\BibitemOpen
  \bibfield  {author} {\bibinfo {author} {\bibfnamefont {T.~E.}\ \bibnamefont
  {Chupp}}, \bibinfo {author} {\bibfnamefont {P.}~\bibnamefont {Fierlinger}},
  \bibinfo {author} {\bibfnamefont {M.~J.}\ \bibnamefont {Ramsey-Musolf}},\
  and\ \bibinfo {author} {\bibfnamefont {J.~T.}\ \bibnamefont {Singh}},\
  }\bibfield  {title} {\bibinfo {title} {Electric dipole moments of atoms,
  molecules, nuclei, and particles},\ }\href
  {https://doi.org/10.1103/RevModPhys.91.015001} {\bibfield  {journal}
  {\bibinfo  {journal} {Rev. Mod. Phys.}\ }\textbf {\bibinfo {volume} {91}},\
  \bibinfo {pages} {015001} (\bibinfo {year} {2019})}\BibitemShut {NoStop}\bibitem [{\citenamefont {Yamaguchi}\ and\ \citenamefont
  {Yamanaka}(2020)}]{muonEDM_limits_small2020}\BibitemOpen
  \bibfield  {author} {\bibinfo {author} {\bibfnamefont {Y.}~\bibnamefont
  {Yamaguchi}}\ and\ \bibinfo {author} {\bibfnamefont {N.}~\bibnamefont
  {Yamanaka}},\ }\bibfield  {title} {\bibinfo {title} {Large long-distance
  contributions to the electric dipole moments of charged leptons in the
  {Standard Model}},\ }\href {https://doi.org/10.1103/PhysRevLett.125.241802}
  {\bibfield  {journal} {\bibinfo  {journal} {Phys. Rev. Lett.}\ }\textbf
  {\bibinfo {volume} {125}},\ \bibinfo {pages} {241802} (\bibinfo {year}
  {2020})}\BibitemShut {NoStop}\bibitem [{\citenamefont {Chivukula}\ \emph {et~al.}(1987)\citenamefont
  {Chivukula}, \citenamefont {Georgi},\ and\ \citenamefont
  {Randall}}]{Chivukula:1987fw}\BibitemOpen
  \bibfield  {author} {\bibinfo {author} {\bibfnamefont {R.~S.}\ \bibnamefont
  {Chivukula}}, \bibinfo {author} {\bibfnamefont {H.}~\bibnamefont {Georgi}},\
  and\ \bibinfo {author} {\bibfnamefont {L.}~\bibnamefont {Randall}},\
  }\bibfield  {title} {\bibinfo {title} {{A Composite Technicolor Standard
  Model of Quarks}},\ }\href {https://doi.org/10.1016/0550-3213(87)90638-9}
  {\bibfield  {journal} {\bibinfo  {journal} {Nucl. Phys. B}\ }\textbf
  {\bibinfo {volume} {292}},\ \bibinfo {pages} {93} (\bibinfo {year}
  {1987})}\BibitemShut {NoStop}\bibitem [{\citenamefont {Hall}\ and\ \citenamefont
  {Randall}(1990)}]{Hall:1990ac}\BibitemOpen
  \bibfield  {author} {\bibinfo {author} {\bibfnamefont {L.~J.}\ \bibnamefont
  {Hall}}\ and\ \bibinfo {author} {\bibfnamefont {L.}~\bibnamefont {Randall}},\
  }\bibfield  {title} {\bibinfo {title} {{Weak scale effective
  supersymmetry}},\ }\href {https://doi.org/10.1103/PhysRevLett.65.2939}
  {\bibfield  {journal} {\bibinfo  {journal} {Phys. Rev. Lett.}\ }\textbf
  {\bibinfo {volume} {65}},\ \bibinfo {pages} {2939} (\bibinfo {year}
  {1990})}\BibitemShut {NoStop}\bibitem [{\citenamefont {Buras}\ \emph {et~al.}(2001)\citenamefont {Buras},
  \citenamefont {Gambino}, \citenamefont {Gorbahn}, \citenamefont {Jager},\
  and\ \citenamefont {Silvestrini}}]{Buras:2000dm}\BibitemOpen
  \bibfield  {author} {\bibinfo {author} {\bibfnamefont {A.~J.}\ \bibnamefont
  {Buras}}, \bibinfo {author} {\bibfnamefont {P.}~\bibnamefont {Gambino}},
  \bibinfo {author} {\bibfnamefont {M.}~\bibnamefont {Gorbahn}}, \bibinfo
  {author} {\bibfnamefont {S.}~\bibnamefont {Jager}},\ and\ \bibinfo {author}
  {\bibfnamefont {L.}~\bibnamefont {Silvestrini}},\ }\bibfield  {title}
  {\bibinfo {title} {{Universal unitarity triangle and physics beyond the
  standard model}},\ }\href {https://doi.org/10.1016/S0370-2693(01)00061-2}
  {\bibfield  {journal} {\bibinfo  {journal} {Phys. Lett. B}\ }\textbf
  {\bibinfo {volume} {500}},\ \bibinfo {pages} {161} (\bibinfo {year}
  {2001})}\BibitemShut {NoStop}\bibitem [{\citenamefont {D'Ambrosio}\ \emph {et~al.}(2002)\citenamefont
  {D'Ambrosio}, \citenamefont {Giudice}, \citenamefont {Isidori},\ and\
  \citenamefont {Strumia}}]{DAmbrosio:2002vsn}\BibitemOpen
  \bibfield  {author} {\bibinfo {author} {\bibfnamefont {G.}~\bibnamefont
  {D'Ambrosio}}, \bibinfo {author} {\bibfnamefont {G.~F.}\ \bibnamefont
  {Giudice}}, \bibinfo {author} {\bibfnamefont {G.}~\bibnamefont {Isidori}},\
  and\ \bibinfo {author} {\bibfnamefont {A.}~\bibnamefont {Strumia}},\
  }\bibfield  {title} {\bibinfo {title} {{Minimal flavor violation: An
  Effective field theory approach}},\ }\href
  {https://doi.org/10.1016/S0550-3213(02)00836-2} {\bibfield  {journal}
  {\bibinfo  {journal} {Nucl. Phys. B}\ }\textbf {\bibinfo {volume} {645}},\
  \bibinfo {pages} {155} (\bibinfo {year} {2002})}\BibitemShut {NoStop}\bibitem [{\citenamefont {Ema}\ \emph {et~al.}(2022)\citenamefont {Ema},
  \citenamefont {Gao},\ and\ \citenamefont {Pospelov}}]{Ema2022indirect}\BibitemOpen
  \bibfield  {author} {\bibinfo {author} {\bibfnamefont {Y.}~\bibnamefont
  {Ema}}, \bibinfo {author} {\bibfnamefont {T.}~\bibnamefont {Gao}},\ and\
  \bibinfo {author} {\bibfnamefont {M.}~\bibnamefont {Pospelov}},\ }\bibfield
  {title} {\bibinfo {title} {Improved indirect limits on muon electric dipole
  moment},\ }\href {https://doi.org/10.1103/PhysRevLett.128.131803} {\bibfield
  {journal} {\bibinfo  {journal} {Phys. Rev. Lett.}\ }\textbf {\bibinfo
  {volume} {128}},\ \bibinfo {pages} {131803} (\bibinfo {year}
  {2022})}\BibitemShut {NoStop}\bibitem [{\citenamefont {Bennett}\ \emph {et~al.}(2009)\citenamefont {Bennett}
  \emph {et~al.}}]{BNLEDM2009}\BibitemOpen
  \bibfield  {author} {\bibinfo {author} {\bibfnamefont {G.~W.}\ \bibnamefont
  {Bennett}} \emph {et~al.} (\bibinfo {collaboration} {Muon $g-2$
  Collaboration}),\ }\bibfield  {title} {\bibinfo {title} {Improved limit on
  the muon electric dipole moment},\ }\href
  {https://doi.org/10.1103/PhysRevD.80.052008} {\bibfield  {journal} {\bibinfo
  {journal} {Phys. Rev. D}\ }\textbf {\bibinfo {volume} {80}},\ \bibinfo
  {pages} {052008} (\bibinfo {year} {2009})}\BibitemShut {NoStop}\bibitem [{\citenamefont {Khaw}\ \emph {et~al.}(2023)\citenamefont {Khaw},
  \citenamefont {Nakai}, \citenamefont {Sato}, \citenamefont {Shigekami},\ and\
  \citenamefont {Zhang}}]{DMedm_2023}\BibitemOpen
  \bibfield  {author} {\bibinfo {author} {\bibfnamefont {K.~S.}\ \bibnamefont
  {Khaw}}, \bibinfo {author} {\bibfnamefont {Y.}~\bibnamefont {Nakai}},
  \bibinfo {author} {\bibfnamefont {R.}~\bibnamefont {Sato}}, \bibinfo {author}
  {\bibfnamefont {Y.}~\bibnamefont {Shigekami}},\ and\ \bibinfo {author}
  {\bibfnamefont {Z.}~\bibnamefont {Zhang}},\ }\bibfield  {title} {\bibinfo
  {title} {A large muon {EDM} from dark matter},\ }\href
  {https://doi.org/10.1007/jhep02(2023)234} {\bibfield  {journal} {\bibinfo
  {journal} {Journal of High Energy Physics}\ }\textbf {\bibinfo {volume}
  {2023}},\ \bibinfo {pages} {234} (\bibinfo {year} {2023})}\BibitemShut
  {NoStop}\bibitem [{\citenamefont {Dekens}\ \emph {et~al.}(2019)\citenamefont {Dekens},
  \citenamefont {de~Vries}, \citenamefont {Jung},\ and\ \citenamefont
  {Vos}}]{Dekens_LQ_2019}\BibitemOpen
  \bibfield  {author} {\bibinfo {author} {\bibfnamefont {W.}~\bibnamefont
  {Dekens}}, \bibinfo {author} {\bibfnamefont {J.}~\bibnamefont {de~Vries}},
  \bibinfo {author} {\bibfnamefont {M.}~\bibnamefont {Jung}},\ and\ \bibinfo
  {author} {\bibfnamefont {K.~K.}\ \bibnamefont {Vos}},\ }\bibfield  {title}
  {\bibinfo {title} {The phenomenology of electric dipole moments in models of
  scalar leptoquarks},\ }\href {https://doi.org/10.1007/jhep01(2019)069}
  {\bibfield  {journal} {\bibinfo  {journal} {J. High Energ. Phys.}\ }\textbf
  {\bibinfo {volume} {2019}}\bibinfo  {number} { (69)}}\BibitemShut {NoStop}\bibitem [{\citenamefont {King}\ \emph {et~al.}(2022)\citenamefont {King} \emph
  {et~al.}}]{FNAL_tracker_2022}\BibitemOpen
\bibfield  {number} {  }\bibfield  {author} {\bibinfo {author} {\bibfnamefont
  {B.~T.}\ \bibnamefont {King}} \emph {et~al.},\ }\bibfield  {title} {\bibinfo
  {title} {{The straw tracking detector for the Fermilab Muon \ensuremath{g-2}
  Experiment}},\ }\href {https://doi.org/10.1088/1748-0221/17/02/P02035}
  {\bibfield  {journal} {\bibinfo  {journal} {{J. Instrum.}}\ }\textbf
  {\bibinfo {volume} {17}},\ \bibinfo {pages} {P02035} (\bibinfo {year}
  {2022})}\BibitemShut {NoStop}\bibitem [{\citenamefont {Aguillard}\ \emph {et~al.}(2023)\citenamefont
  {Aguillard} \emph {et~al.}}]{FNAL_PRL_2023}\BibitemOpen
  \bibfield  {author} {\bibinfo {author} {\bibfnamefont {D.}~\bibnamefont
  {Aguillard}} \emph {et~al.} (\bibinfo {collaboration} {Muon $g-2$
  Collaboration}),\ }\bibfield  {title} {\bibinfo {title} {Measurement of the
  positive muon anomalous magnetic moment to 0.20 ppm},\ }\href
  {https://doi.org/10.1103/physrevlett.131.161802} {\bibfield  {journal}
  {\bibinfo  {journal} {Phys. Rev. Lett.}\ }\textbf {\bibinfo {volume} {131}},\
  \bibinfo {pages} {161802} (\bibinfo {year} {2023})}\BibitemShut {NoStop}\bibitem [{\citenamefont {Grange}\ \emph {et~al.}(2018)\citenamefont {Grange}
  \emph {et~al.}}]{FNAL_TDR2018}\BibitemOpen
  \bibfield  {author} {\bibinfo {author} {\bibfnamefont {J.}~\bibnamefont
  {Grange}} \emph {et~al.} (\bibinfo {collaboration} {Muon $g-2$
  Collaboration}),\ }\href {https://arxiv.org/abs/1501.06858} {\bibinfo {title}
  {Muon $g-2$ technical design report}} (\bibinfo {year} {2018}),\ \Eprint
  {https://arxiv.org/abs/1501.06858} {arXiv:1501.06858} \BibitemShut {NoStop}\bibitem [{\citenamefont {Stratakis}\ \emph {et~al.}(2017)\citenamefont
  {Stratakis} \emph {et~al.}}]{FNAL_accel_2017}\BibitemOpen
  \bibfield  {author} {\bibinfo {author} {\bibfnamefont {D.}~\bibnamefont
  {Stratakis}} \emph {et~al.},\ }\bibfield  {title} {\bibinfo {title}
  {{Accelerator performance analysis of the Fermilab Muon Campus}},\ }\href
  {https://doi.org/10.1103/PhysRevAccelBeams.20.111003} {\bibfield  {journal}
  {\bibinfo  {journal} {Phys. Rev. Accel. Beams}\ }\textbf {\bibinfo {volume}
  {20}},\ \bibinfo {pages} {111003} (\bibinfo {year} {2017})}\BibitemShut
  {NoStop}\bibitem [{\citenamefont {Khaw}\ \emph {et~al.}(2019)\citenamefont {Khaw} \emph
  {et~al.}}]{FNAL_calos2019}\BibitemOpen
  \bibfield  {author} {\bibinfo {author} {\bibfnamefont {K.}~\bibnamefont
  {Khaw}} \emph {et~al.},\ }\bibfield  {title} {\bibinfo {title} {Performance
  of the muon $g-2$ calorimeter and readout systems measured with test beam
  data},\ }\href {https://doi.org/https://doi.org/10.1016/j.nima.2019.162558}
  {\bibfield  {journal} {\bibinfo  {journal} {Nucl. Instrum. Meth. A}\ }\textbf
  {\bibinfo {volume} {945}},\ \bibinfo {pages} {162558} (\bibinfo {year}
  {2019})}\BibitemShut {NoStop}\bibitem [{\citenamefont {Abi}\ \emph {et~al.}(2021)\citenamefont {Abi} \emph
  {et~al.}}]{FNAL_PRL_2021}\BibitemOpen
  \bibfield  {author} {\bibinfo {author} {\bibfnamefont {B.}~\bibnamefont
  {Abi}} \emph {et~al.} (\bibinfo {collaboration} {Muon $g-2$ Collaboration}),\
  }\bibfield  {title} {\bibinfo {title} {Measurement of the positive muon
  anomalous magnetic moment to 0.46 ppm},\ }\href
  {https://doi.org/10.1103/PhysRevLett.126.141801} {\bibfield  {journal}
  {\bibinfo  {journal} {Phys. Rev. Lett.}\ }\textbf {\bibinfo {volume} {126}},\
  \bibinfo {pages} {141801} (\bibinfo {year} {2021})}\BibitemShut {NoStop}\bibitem [{\citenamefont {Schreckenberger}\ \emph {et~al.}(2021)\citenamefont
  {Schreckenberger} \emph {et~al.}}]{kickerpaper}\BibitemOpen
  \bibfield  {author} {\bibinfo {author} {\bibfnamefont {A.~P.}\ \bibnamefont
  {Schreckenberger}} \emph {et~al.},\ }\bibfield  {title} {\bibinfo {title}
  {{The fast non-ferric kicker system for the Muon g\ensuremath{-}2 Experiment
  at Fermilab}},\ }\href {https://doi.org/10.1016/j.nima.2021.165597}
  {\bibfield  {journal} {\bibinfo  {journal} {Nucl. Instrum. Methods Phys.
  Res., Sect. A}\ }\textbf {\bibinfo {volume} {1011}},\ \bibinfo {pages}
  {165597} (\bibinfo {year} {2021})}\BibitemShut {NoStop}\bibitem [{\citenamefont {Albahri}\ \emph {et~al.}(2021)\citenamefont {Albahri}
  \emph {et~al.}}]{FNAL_BeamDyn}\BibitemOpen
  \bibfield  {author} {\bibinfo {author} {\bibfnamefont {T.}~\bibnamefont
  {Albahri}} \emph {et~al.} (\bibinfo {collaboration} {Muon $g-2$
  Collaboration}),\ }\bibfield  {title} {\bibinfo {title} {{Beam dynamics
  corrections to the Run-1 measurement of the muon anomalous magnetic moment at
  Fermilab}},\ }\href {https://doi.org/10.1103/PhysRevAccelBeams.24.044002}
  {\bibfield  {journal} {\bibinfo  {journal} {Phys. Rev. Accel. Beams}\
  }\textbf {\bibinfo {volume} {24}},\ \bibinfo {pages} {044002} (\bibinfo
  {year} {2021})}\BibitemShut {NoStop}\bibitem [{\citenamefont {Aguillard}\ \emph {et~al.}(2024)\citenamefont
  {Aguillard} \emph {et~al.}}]{FNAL_PRD_2024}\BibitemOpen
  \bibfield  {author} {\bibinfo {author} {\bibfnamefont {D.~P.}\ \bibnamefont
  {Aguillard}} \emph {et~al.} (\bibinfo {collaboration} {Muon
  $g\ensuremath{-}2$ Collaboration}),\ }\bibfield  {title} {\bibinfo {title}
  {Detailed report on the measurement of the positive muon anomalous magnetic
  moment to 0.20 ppm},\ }\href {https://doi.org/10.1103/PhysRevD.110.032009}
  {\bibfield  {journal} {\bibinfo  {journal} {Phys. Rev. D}\ }\textbf {\bibinfo
  {volume} {110}},\ \bibinfo {pages} {032009} (\bibinfo {year}
  {2024})}\BibitemShut {NoStop}\bibitem [{\citenamefont {Aguillard}\ \emph {et~al.}(2025)\citenamefont
  {Aguillard} \emph {et~al.}}]{FNAL_PRL_2025}\BibitemOpen
  \bibfield  {author} {\bibinfo {author} {\bibfnamefont {D.~P.}\ \bibnamefont
  {Aguillard}} \emph {et~al.} (\bibinfo {collaboration} {Muon
  $g\ensuremath{-}2$ Collaboration}),\ }\bibfield  {title} {\bibinfo {title}
  {Measurement of the positive muon anomalous magnetic moment to 127 ppb},\
  }\href {https://doi.org/10.1103/7clf-sm2v} {\bibfield  {journal} {\bibinfo
  {journal} {Phys. Rev. Lett.}\ }\textbf {\bibinfo {volume} {135}},\ \bibinfo
  {pages} {101802} (\bibinfo {year} {2025})}\BibitemShut {NoStop}\bibitem [{\citenamefont {Allison}\ \emph {et~al.}(2016)\citenamefont {Allison}
  \emph {et~al.}}]{Geant42016}\BibitemOpen
  \bibfield  {author} {\bibinfo {author} {\bibfnamefont {J.}~\bibnamefont
  {Allison}} \emph {et~al.},\ }\bibfield  {title} {\bibinfo {title} {Recent
  developments in {Geant4}},\ }\href
  {https://doi.org/https://doi.org/10.1016/j.nima.2016.06.125} {\bibfield
  {journal} {\bibinfo  {journal} {Nuclear Instruments and Methods in Physics
  Research Section A: Accelerators, Spectrometers, Detectors and Associated
  Equipment}\ }\textbf {\bibinfo {volume} {835}},\ \bibinfo {pages} {186}
  (\bibinfo {year} {2016})}\BibitemShut {NoStop}\bibitem [{\citenamefont {Lyons}\ \emph {et~al.}(1988)\citenamefont {Lyons},
  \citenamefont {Gibaut},\ and\ \citenamefont {Clifford}}]{Lyons:1988rp}\BibitemOpen
  \bibfield  {author} {\bibinfo {author} {\bibfnamefont {L.}~\bibnamefont
  {Lyons}}, \bibinfo {author} {\bibfnamefont {D.}~\bibnamefont {Gibaut}},\ and\
  \bibinfo {author} {\bibfnamefont {P.}~\bibnamefont {Clifford}},\ }\bibfield
  {title} {\bibinfo {title} {{How to Combine Correlated Estimates of a Single
  Physical Quantity}},\ }\href {https://doi.org/10.1016/0168-9002(88)90018-6}
  {\bibfield  {journal} {\bibinfo  {journal} {Nucl. Instrum. Meth. A}\ }\textbf
  {\bibinfo {volume} {270}},\ \bibinfo {pages} {110} (\bibinfo {year}
  {1988})}\BibitemShut {NoStop}\bibitem [{\citenamefont {Feldman}\ and\ \citenamefont
  {Cousins}(1998)}]{FeldmanCousins}\BibitemOpen
  \bibfield  {author} {\bibinfo {author} {\bibfnamefont {G.~J.}\ \bibnamefont
  {Feldman}}\ and\ \bibinfo {author} {\bibfnamefont {R.~D.}\ \bibnamefont
  {Cousins}},\ }\bibfield  {title} {\bibinfo {title} {Unified approach to the
  classical statistical analysis of small signals},\ }\href
  {https://doi.org/10.1103/physrevd.57.3873} {\bibfield  {journal} {\bibinfo
  {journal} {Physical Review D}\ }\textbf {\bibinfo {volume} {57}},\ \bibinfo
  {pages} {3873–3889} (\bibinfo {year} {1998})}\BibitemShut {NoStop}\end{thebibliography}
\end{document}